\documentclass[12pt]{article}

\usepackage{amsmath}

\usepackage{amsfonts}

\usepackage{amssymb}

\usepackage{hyperref}

\usepackage[latin1]{inputenc}

\usepackage{graphics}

\usepackage{graphicx}

\usepackage{epsfig}

\usepackage{amsthm}

\usepackage{mathrsfs}

\usepackage{latexsym}

\usepackage{lineno}

\usepackage{latexsym}

\usepackage{float}

\usepackage{mathtools}

\usepackage[T1]{fontenc}

\usepackage{ae}

\usepackage{authblk}

\modulolinenumbers[5]

\newenvironment{destaque}{\begin{quotation}\small}{\end{quotation}}
\usepackage[figurename=Fig.]{caption}
\usepackage{subcaption}
\newcommand{\h}{\hspace{.5cm}}

\newcommand{\mvl}[1] {\left\langle #1 \right\rangle}
\newcommand{\ket}[1] {\lvert #1 \rangle}

\newcommand{\bkt}[2] {\left\langle #1 \mid #2 \right\rangle}

\date{}

\begin{document}

%%%%%%%%%%%%%%%%%%%%%%%%%%%%%%%%%%%%%%%%%%%%%%%%%%%%%%%%%%%%%%%%%%%%%%%%%%%%%%%%%%%%%%%%%%%%%%%%%%%
%%%%%%%%%%%%%%%%%%%%%%%%%%%%%%%%%% TITLE and AUTHORS %%%%%%%%%%%%%%%%%%%%%%%%%%%%%%%%%%%%%%%%%%%%%%
%%%%%%%%%%%%%%%%%%%%%%%%%%%%%%%%%%%%%%%%%%%%%%%%%%%%%%%%%%%%%%%%%%%%%%%%%%%%%%%%%%%%%%%%%%%%%%%%%%%
\title{\bf Quantum Cosmology with Dynamical Vacuum in a Minimal-Length Scenario}

%%%%%%%%%%%%%%%%%%%%%%%%%%%%%%%%%%%%%%%%%%%%%%%%%%%%%%%%%%%%%%%%%%%%%%%%%%%%%%%%%%%%%%%%%%%%%%%%%%%%%%%%%%%%%%%% AUTHORS %%%%%%%%%%%%%%%%%%%%%%%%%%%%%%%%%%%%%%%%%%%%%%%%%%%%%%%

\author{M. F. Gusson, A. Oakes O. Gon\c{c}alves\thanks{Actual adress: Instituto Federal Baiano, Campus Itapetinga, BA.}, R. G. Furtado,\\
 J. C. Fabris\thanks{National Research Nuclear University MEPhI, Kashirskoe sh. 31, Moscow 115409, Russia.} \thanks{julio.fabris@cosmo-ufes.org} ~and J. A. Nogueira\thanks{jose.nogueira@ufes.br}}
\affil{Universidade Federal do Esp\'{\i}rito Santo -- Ufes\\ Vit\' oria, Esp\'{\i}rito Santo, 29.075-910, Brasil}

%%%%%%%%%%%%%%%%%%%%%%%%%%%%%%%%%%%%%%%%%%%%%%%%%%%%%%%%%%%%%%%%%%%%%%%%%%%%%%%%%%%%%%%%%%%%%%%%%%%%%%%%%%%%%%%%%%%%%%%%%%%%%%%%%%%%%%%%%%%%%%%%%%%%%%%%%%%%%%%%%%%%%%%%%%%%%%%%

\maketitle
%%%%%%%%%%%%%%%%%%%%%%%%%%%%%%%%%%%%%%%%%%%%%%%%%%%%%%%%%%%%%%%%%%%%%%%%%%%%%%%%%%%%%%%%%%%%%%%%%%%

%%%%%%%%%%%%%%%%%%%%%%%%%%%%%%%%%%%%%%%%%%%%%%%%%%%%%%%%%%%%%%%%%%%%%%%%%%%%%%%%%%%%%%%%%%%%%%%%%%%
%%%%%%%%%%%%%%%%%%%%%%%%%%%%%%%%%% ABSTRACT %%%%%%%%%%%%%%%%%%%%%%%%%%%%%%%%%%%%%%%%%%%%%%%%%%%%%%%
%%%%%%%%%%%%%%%%%%%%%%%%%%%%%%%%%%%%%%%%%%%%%%%%%%%%%%%%%%%%%%%%%%%%%%%%%%%%%%%%%%%%%%%%%%%%%%%%%%%
\begin{abstract}
\begin{destaque}
 In this work, we consider effects of the dynamical vacuum in quantum cosmology in presence of a minimum length introduced by the GUP (generalized uncertainty principle) related to the modified commutation relation $[\hat{X},\hat{P}] := \frac{i\hbar}{ 1 - \beta\hat{P}^2 }$ . We determine the wave function of the Universe $ \psi_{qp}(\xi,t)$, which is solution of the modified Wheeler-DeWitt equation in the representation of the quasi-position space, in the limit where the scale factor of the Universe is small. Although $\psi_{qp}(\xi,t)$ is a physically acceptable state it is not a realizable state of the Universe because $ \psi_{qp}(\xi,t)$ has infinite norm, as in the ordinary case with no minimal length.\\
{\scriptsize PACS numbers: 98.80.Qc, 04.60.-m, 04.50.Kd,  }\\
{\scriptsize Keywords: Minimal length; generalized uncertainty principle; quantum gravity; quantum cosmology; dynamical vacuum.}
\end{destaque}
\end{abstract}
%%%%%%%%%%%%%%%%%%%%%%%%%%%%%%%%%%%%%%%%%%%%%%%%%%%%%%%%%%%%%%%%%%%%%%%%%%%%%%%%%%%%%%%%%%%%%%%%%%%
%%%%%%%%%%%%%%%%%%%%%%%%%%%%%%%%%% END Abstract %%%%%%%%%%%%%%%%%%%%%%%%%%%%%%%%%%%%%%%%%%%%%%%%%%%
%%%%%%%%%%%%%%%%%%%%%%%%%%%%%%%%%%%%%%%%%%%%%%%%%%%%%%%%%%%%%%%%%%%%%%%%%%%%%%%%%%%%%%%%%%%%%%%%%%%

% \linenumbers
\pagenumbering{arabic}

%%%%%%%%%%%%%%%%%%%%%%%%%%%%%%%%%%%%%%%%%%%%%%%%%%%%%%%%%%%%%%%%%%%%%%%%%%%%%%%%%%%%%%%%%%%%%%%%%%%%%%%%%%%%%%%%%%%%%%%%%%%%%%%%%%%%%%%%%%%%%%%%%%%%%%%%%%%%

%%%%%%%%%%%%%%%%%%%%%%%%%%%%%%%%%%%%%%%%%%%%%%%%%%%%%%%%%%%%%%%%%%%%%%%%%%%%%%
%%%%%%%%%%%%%%%%%%%%%%  INTRODUCTION  %%%%%%%%%%%%%%%%%%%%%%%%%%%%%%%%%%%%%%%%
%%%%%%%%%%%%%%%%%%%%%%%%%%%%%%%%%%%%%%%%%%%%%%%%%%%%%%%%%%%%%%%%%%%%%%%%%%%%%%
\section{Introduction}
\label{introd}
%%%%%%%%%%%%%%%%%%%%%%%%%%%%%%%%%%%%%%%%

\h The Standard Cosmological Model (SCM) is based on an expanding universe which has a very hot and dense origin. The extrapolation of the SCM for very early times leads to energy scales deep in the Planck regime and, strictly speaking, the expanding universe would have a singular initial state. On the other hand, in order to explain many features of the observed Universe today, a period of short but rapid, almost exponential, expansion, which is called {\it inflationary phase}, is necessary in the very early Universe.
It is expected that in the primordial Universe quantum effects are relevant and a quantum approach is necessary to describe the gravitational effects. A quantum approach to gravity is also necessary in the study of black holes. The features of compact astrophysical objects like neutron stars, and even white dwarfs, may be also affected by quantum gravity effects. However, we have no quantum gravity theory which is fully acceptable, although there are many proposals for it.
An important aspect of almost all proposals for quantization of the gravity is the prediction of the existence of a minimal length. Thus, an effective description of the effects of quantum gravity through a phenomenological approach can be obtained by considering a minimal-length scenario \cite{Hossenfelder:2006,Kober1:2012,Nouicer1:2012,Ong2:2018}. Some results in the literature indicate that the introduction of a minimal length could also solve the problems like the initial singularity (big bang singularity) \cite{Vakili:2007,Battisti1:2007,Battisti2:2007,Battisti1:2008,Battisti2:2008,Ali:2014,Faizal:2014,Faizal:2016,Aydiner1:2020}, the black hole complete evaporation (catastrophic evaporation of a black hole) \cite{Nouicer1:2012,Nouicer1:2007,Nouicer2:2007,Kim:2008,Nouicer:2009,Ong1:2018}, the cosmological constant \cite{Faizal:2016,Faizal:2015,Ong3:2018,Tawfik1:2020}, the Chandrasekhar limit \cite{Ong2:2018,Ong1:2018,Ong3:2018}, the instability of the Einstein static universe (emergent universe) \cite{Nozari1:2018}, the vanishing of cosmological constant in multiverse theories \cite{Faizal:2015}, and so on.

We can obtain a minimal-length scenario by  modifying the Heisenberg uncertainty principle (HUP) between the position and momentum operators. There are several different proposals of modification of the HUP which introduce a minimal-length scenario. Those modified uncertainty principles are called Generalized Uncertainty Principle (GUP)\footnote{Readers interested in finding out more about minimal-length literature may consult References \cite{Nouicer:2009,Takeuchi:2011,Hossenfelder:2013,Tawfik1:2014,Tawfik2:2015}.}.
GUP's have been derived from different contexts such as string theory \cite{Veneziano:1986,Amati:1987,Amati:1989,Gross:1987,Gross:1988,konishi:1990}, black hole physics  \cite{Faizal:2015,Maggiore1:1993,Scardigli:1999,Park1:2008} and extra dimensions \cite{Scardigli:2003}.

Although the most common modified or deformed commutation relation (MCR) associated with a GUP (KMM GUP \cite{Kempf1:1995}) is quadratic in the momentum operator, effects of a MCR with a linear term in the momentum operator, which leads to a minimal length and a maximum momentum \cite{Ali:2009}, have been studied in contexts of cosmological models \cite{Ali:2014,Faizal:2014,Majumder1:2011,Majumder2:2011,Majumder3:2011}. Nevertheless, apparently this GUP leads to non-unitary theories \cite{Majumder2:2011} and it is not compatible with current data available \cite{Salzano1:2020}.

In general, the inflationary phase is described by introducing a scalar field. Hence, in an inflationary quantum cosmology theory  it is necessary to perform the second quantization of that scalar field. Consequently, in a minimal-length scenario that scalar field and its conjugate momentum have to obey the modified commutation relations \cite{Kempf1:2001,Hassan:2003,Kempf:2005,Kempf:2006}.
As it is well known, the ordinary approach of the third quantization\footnote{The 3rd quantization consists of initially considering the Wheeler-DeWitt equation as a classical equation of a classical field (the wave function of the Universe in the 2nd quantization) which then is turned an operator. In this way universes can be created and annihilated in a multiverses theory.} of the Wheeler-DeWitt (WDW) equation leads to the vanishing of the cosmological constant. However, it might be possible to obtain a non-zero cosmological constant in a minimal-length scenario \cite{Faizal:2015}.

Many authors have studied effects of a minimal length in cosmology using a classical approach in which the Poisson brackets are modified according to the correspondence principle \cite{Ali:2014,Nozari1:2018,Vakili:2009,Tawfik1:2015,Faizal1:2017,Darabi1:2017,Moumni1:2020}.   
In the vast majority of the GUP formulation the parameter related to the minimal length (deformation parameter) is positive. Although GUP with positive deformation parameter prevents black holes to evaporate completely, at the same time they remove the Chandrasekhar limit\footnote{The maximum mass of a stable white dwarf star.}, that is, a white dwarf star could turn out arbitrarily large \cite{Moussa1:2015,Rashidi1:2016}. In \cite{Ong1:2018,Ong3:2018} the author shows that GUP with a negative deformation parameter can restore the Chandrasekhar limit and, in spite of allowing a black hole to evaporate completely, this evaporation takes an infinite amount of time. Even though a negative deformation parameter is unusual, it is consistent with a description in which the universe has an underlying crystal lattice-like structure \cite{Scardigli:2010}.

Mu-In Park proposed that the HUP can be modified to include the cosmology constant term, which is called extend uncertainty principle (EUP) \cite{Park1:2008,Andrade1:2016}. Whereas GUP has a quadratic term in the momentum uncertainty, EUP has a quadratic term in the position uncertainty. GUP that has also a quadratic term in the position uncertainty is called GEUP. In \cite{Ong2:2018} the authors use GEUP to show that a non-zero cosmological constant can restore the Chandrasekhar limit. Effects of GEUP in cosmology has been studied in \cite{Kouwn1:2018}.
Modified uncertainty principles that induce a maximum length have been employed in order to describe the cosmological particle horizon \cite{Perivolaropoulos1:2017}. This modified uncertainty principle has terms proportional to even powers in the position uncertainty \cite{Perivolaropoulos1:2019}.
Cosmological observational data have been used to obtain constraints for GUP parameters \cite{Tawfik1:2020,Salzano1:2020,Tawfik1:2015,Kouwn1:2018,Nozari1:2014}.

Even though the de Sitter Cosmology Model describes the rapidly expansion phase of the Universe, during which vacuum energy dominates, its traditional treatment considers a system without physical content since there is only  a single degree of freedom and one constraint \cite{Vilenkin:1994}. Schutz formalism \cite{Schutz1:1970,Schutz2:1970}, which describes a relativistic fluid interacting with the gravity field, can be used to regard the vacuum as a dynamic entity having different degrees of freedom. This procedure overcomes the difficulties found in de Sitter Model and it leads to a natural way  of introducing a variable playing the role of time. Then, the Dynamical Vacuum Model can be implemented in a homogeneous and isotropic universe filled with vacuum fluid whose state equation is $P = - \rho$, which is classically treated according to the Schutz canonical formalism. This process implies a linear term in one of the momenta in the Hamiltonian whose associated degree of freedom will play the role of time \cite{Flavio:1998}.

In general, with the intention of avoiding the problems caused by the WDW equation defined in the superspace\footnote{As far as we know, there is no known general solution for the WDW equation in the superspace.} (the space of all possible three dimensional metrics), the process of quantization is performed using the mini-superspace  approach \cite{DeWitt:1967}, where an infinite number of degrees of freedom of the gravity field is frozen\footnote{In fact, the mini-superspace approach is an approximation in which are only considered the largest gravity field wavelength modes of the order of the size of the Universe.} and remaining degrees of freedom are turned operators.

Of course, if GUP is a fundamental aspect of the nature then not only the position and the momentum have to obey the GUP but also every other variables which will be quantized \cite{Pedram1:2015}. Since GUP corresponds to a modification of the commutation relation between the operator and its conjugate operator, the WDW equation in a minimal-length scenario can be obtained imposing that some or all of those operators (which have come from the quantization of remaining degrees of freedom in the mini-superspace approach) and its conjugate momentum operators satisfy modified commutation relations.

In this work, our primary purpose is to determine the corrections in the wave function of the Universe due to the use of a specific generalized uncertainty principle (GUP) in a quantum cosmology model. Hence, we determine the modified WDW equation up to ${\cal O}(\beta^{2})$ considering a quantum cosmology model of dynamical vacuum in a minimal-length scenario induced by the commutation relation proposed by P. Pedram \cite{Pedram1:2012,Pedram2:2012},
\begin{equation*}
	[\hat{X},\hat{P}] := \frac{i\hbar}{1 - \beta\hat{P}^2},
\end{equation*}
which induces a minimal uncertainty in the position, $\Delta x_{min} = \frac{3 \sqrt{3}}{4} \hbar \sqrt{\beta}$, and an upper bound on the conjugate momentum, $P_{max} = \frac{1}{\sqrt{\beta}}$. In the equation above $\beta$ is a parameter related to the minimal length. An important aspect of the GUP associated with this modified commutation relation is that it is not perturbative (in the minimal length). Moreover, it is consistent with Doubly Special Relativity (DSR) theories which predicts an upper bound for particle momentum \cite{Nozari1:2014,Camelia1:2002,Camelia2:2002,Magueijo1:2002} and it is in agreement with several theoretical proposals for quantum gravity \cite{Pedram1:2012}. We find the wave function of the Universe for small scale factors in the formal representation of the ``position'' space, in fact representation of the scale-factor space\footnote{We say representation of the ``position'' space keeping in mind that it should be the more correct to call represetation of the scale-factor space.}. However, it is not possible to obtain from wave function any physical information since the eigenstates of the ``position'' operators are not physical states \cite{Kempf1:1995}. We overcome this problem by obtaining the wave function in the representation of the quasi-position space as a superposition of wave functions in the formal representation of the ``position'' space \cite{Pedram:2013}. Even if it may be questioned whether this procedure is correct, we find the modified WDW equation in the representation of the quasi-position space and we show that the wave function in the quasi-position space previously obtained is solution this equation, as we expected. Although the wave functions in the quasi-position space, which are eigenfunctions of the cosmological constant, are physically acceptable states they are not realizable states of the Universe because have infinite norm. Consequently, it will be necessary to construct wave packets from the superposition of eigenfunctions with different eigenvalues of the cosmological constant. 
Last but not least, it should be said that the quantization process occurs into the mini-superspace scenario in which the scale factor is the only degree of freedom. As far as we know, the first applications of a minimal-length scenario to a mini-superspace dynamics can be found in \cite{Vakili:2007,Battisti2:2007,Battisti1:2008}.

In short, in this paper we intend to study a dynamical vacuum, using a description given by the Schutz formalism, in a quantum cosmological scenario where the minimal-length proposal is explicitly considered. Our main aim is to verify how GUP affects the solutions for the WDW equation with a dynamical vacuum state, in the mini-superspace, in comparison with the usual approach.

The paper is organized in the following way. In Section \ref{MLS}, we describe the minimal-length scenario used, presenting its main results. In Section \ref{Model}, we describe the ordinary cosmology model, that is, in a scenario without minimal length. We obtain the gravity action and the fluid action according to the Schutz formalism. In Section \ref{QCML}, we determine the modified WDW equation in the formal representation of the ``position'' space and we find its solution for small values of the scale factor. We also discuss about the validity range for which the approach employed here is consistent. Then, we find the physically acceptable solutions, that is, the wave function of the Universe in the representation of the quasi-position space. Last we show that the norm of that wave function is infinite, as in the ordinary case with no minimal length. In Section \ref{Concl}, we present our comments and conclusions.

%%%%%%%%%%%%%%%%%%%%%%%%%%%%%%%%%%%%%%%%%%%%%%%%%%%%%%%%%%%%%%%%%%%%%%%%%%%%%%
%%%%%%%%%%%%%%%%%%%%%  END Introduction  %%%%%%%%%%%%%%%%%%%%%%%%%%%%%%%%%%%%%
%%%%%%%%%%%%%%%%%%%%%%%%%%%%%%%%%%%%%%%%%%%%%%%%%%%%%%%%%%%%%%%%%%%%%%%%%%%%%%

%%%%%%%%%%%%%%%%%%%%%%%%%%%%%%%%%%%%%%%%%%%%%%%%%%%%%%%%%%%%%%%%%%%%%%%%%%%%%%
%%%%%%%%%%%%%%%%%%%%%  MINIMAL-LENGTH SCENARIO  %%%%%%%%%%%%%%%%%%%%%%%%%%%%%%
%%%%%%%%%%%%%%%%%%%%%%%%%%%%%%%%%%%%%%%%%%%%%%%%%%%%%%%%%%%%%%%%%%%%%%%%%%%%%%
\section{Minimal-Length Scenario}
\label{MLS}
%%%%%%%%%%%%%%%%%%%%%%%%%%%%%%%%%%%%%%%%%%%%%%%%%%%%%%%%%%%%%%%%%%%%%%%%%%%%%%

\h In a quantum approach, a minimal length can be introduced by modifying the HUP in order to implement a non-zero minimal uncertainty in the position\footnote{It is not trivial to show that a non-zero minimal uncertainty in position can be interpreted as a minimal length \cite{Takeuchi:2011,Kempf1:1995,Mead:1964}.}. There are many proposals for modification of HUP \cite{Nouicer1:2007,Kempf1:1995,Ali:2009,Darabi1:2017,Perivolaropoulos1:2017,Pedram1:2012,Pedram2:2012,Chung1:2020}.

In this work we concern with the GUP
\begin{equation}
	\label{gup}
	\Delta X \Delta P \geq \frac{\frac{\hbar}{2}}{ 1 - \beta \left[(\Delta P)^2 + \langle \hat{P}  \rangle^{2}  \right]},
\end{equation}
proposed by P. Pedram \cite{Pedram1:2012,Pedram2:2012}, which induces a non-zero minimal uncertainty in the position given by
\begin{equation}
	\label{minu}
	\Delta x_{min} = \frac{3 \sqrt{3}}{4}\hbar \sqrt{\beta},
\end{equation}
where $\beta$ is a parameter related to the minimal length. We choose the GUP (\ref{gup}) because it is consistent with several proposals for quantum gravity, such as string theory, loop quantum gravity, and it also introduces a maximal measured momentum\footnote{It is worth noting that this GUP does not induce a maximum uncertainty in the conjugate momentum.},
\begin{equation}
	\label{mmon}
	P_{max} = \frac{1}{\sqrt{\beta}},
\end{equation}
which is in agreement with the DSR. Another important aspect this GUP is that it is not perturbative. Consequently, if $\beta$ is small then (\ref{gup}) can be expanded until any order in $\beta$ we wish.

Since, 
\begin{equation}
	\label{cr}
	\Delta X \Delta P \geq \frac{1}{2}  \left| \mvl{[\hat{X},\hat{P}]} \right|  ,
\end{equation}
then related to the GUP (\ref{gup}) we have the modified commutation relation,
\begin{equation}
\label{drc}
	[\hat{X},\hat{P}] := \frac{i\hbar}{1 - \beta\hat{P}^2}.
\end{equation}

Although, the representation of the operators:
\begin{equation}
	\label{xrepx}
	\hat{X} = \hat{x},
\end{equation}
\begin{equation}
	\label{xrepp}
	\hat{P} = \hat{p} + \frac{\beta}{3}\hat{p}^{3} + \frac{\beta^{2}}{3} \hat{p}^{5} +\frac{4\beta^{3}}{9}\hat{p}^{7} + \dots,
\end{equation}
where $\hat{x}$ and $\hat{p}$ are the ordinary operators of position and momentum satisfying the canonical commutation relation $ [\hat{x},\hat{p}] := i\hbar $, is not an exact representation of the algebra (\ref{drc}), it preserves the ordinary form of the position operator. Thus, in this representation of ``position'' space we have,
\begin{equation}
	\label{repx}
	\langle x | \hat{X} | \psi(t) \rangle = x \psi (x,t),
\end{equation}
$$
\langle x | \hat{P} | \psi(t) \rangle =
$$
\begin{equation}
	\label{repp}
	\left( -i \hbar  \frac{\partial} {\partial x} +  i \hbar^{3} \frac{\beta}{3} \frac{\partial^{3}}{\partial 		x^{3}} - i \hbar^{5} \frac{\beta^{2}}{3} \frac{\partial^{5}}{\partial 			x^{5}} + \dots  \right) \psi (x,t),
\end{equation}
where $\ket{x}$ are the state eigenvectors of the position operator.

This representation of ``position'' space is only formal since the eigenvalues of the position operator (\ref{xrepx}) are not physical states and consequently they do not belong to the Hilbert space. This is because the position operator uncertainty vanishes when it is calculated in any of its eigenstates. But that is physically impossible since $\Delta x \geq \frac{3 \sqrt{3}}{4}\hbar \sqrt{\beta} $ for all physically allowable state in a minimal-length scenario. However, all information on position can be accessible through the maximal localization states, defined as,
\begin{equation}
\label{mls}
	\langle \psi_{\xi}^{ml} | \hat{X} | \psi_{\xi}^{ml} \rangle = \xi
\end{equation}
and
\begin{equation}
	\left( \Delta x \right)_{\ket{\psi_{\xi}^{ml}}} = \left( \Delta x \right)_{min}.
\end{equation}

In the DGS (Detournay, Gabriel and Spindel) approach \cite{Detournay:2002} the maximal localization states are found to be,\footnote{Note that it  is in the representation of the ordinary momentum space.}
\begin{equation}
	\psi_{\xi}^{ml}(p) = \sqrt{\frac{2\sqrt{3}}{2}} \exp \left( -\frac{i}{\hbar} \xi p \right) \cos\left( \frac{3 \pi}{4} \sqrt{\beta} p \right).
\end{equation}

The representation of quasi-position space is obtained by projecting the state vectors onto the  maximal localization states,
\begin{equation}
	\label{qprep}
	\langle \psi_{\xi}^{ml} | \psi(t) \rangle = \psi_{qp}(\xi,t).
\end{equation}

The action of the position and the momentum operators on the quasi-position space are given by,
\begin{equation}
	\label{qprepX}
	\langle \psi_{\xi}^{ml} | \hat{X} | \psi(t) \rangle = \left[ \xi + ib \sqrt{\beta}\tan \left(-ib\sqrt{\beta} \frac{\partial}{\partial \xi} 			\right) \right] \psi_{qp}(\xi,t)
\end{equation}
and
\begin{equation}
	\label{qprepP}
	\langle \psi_{\xi}^{ml} | \hat{P} | \psi(t) \rangle = \langle \psi_{\xi}^{ml} | \left( \hat{p} +  \frac{\beta}{3} \hat{p}^{3} +  \frac{\beta^{2}}		{3} \hat{p}^{5} + \dots \right) | \psi(t) \rangle,
\end{equation}
with
\begin{equation}
	\langle \psi_{\xi}^{ml} | \hat{p} | \psi(t) \rangle = -i \hbar 					\frac{\partial \psi_{qp}(\xi,t)}{\partial \xi},
\end{equation}
and
\begin{equation}
	\label{b}
	b := \frac{3 \pi \hbar}{4}.
\end{equation}

Hence, up to ${\cal O}(\beta^{2})$ we have
\begin{equation}
	\label{qprepX}
	\langle \psi_{\xi}^{ml} | \hat{X} | \psi(t) \rangle = \left( \xi + \beta b^{2} \frac{\partial}{\partial \xi} - \beta^{2} \frac{b^{4}}{3} 				\frac{\partial^{3}}{\partial \xi^{3}} \right) \psi_{qp}(\xi,t),
\end{equation}
and
$$
\langle \psi_{\xi}^{ml} | \hat{P} | \psi(t) \rangle =
$$
\begin{equation}
\label{qprepP}
 \left( -i \hbar \frac{\partial}{\partial \xi} + i \hbar^{3} \frac{\beta}{3} 					\frac{\partial^{3}}{\partial \xi^{3}} - i \hbar^{5} \frac{\beta^{2}}{3} 		\frac{\partial^{5}}{\partial \xi^{5}} \right) \psi_{qp}(\xi,t).
\end{equation}

Finally, one can show that the wave function in the quasi-position space is a superposition of the wave functions in the formal ``position'' space, given by
\begin{equation}
	\label{wfqp}
	\psi_{qp}(\xi) = \frac{1}{\sqrt{2}} \left[ \psi(\xi + b\sqrt{\beta} ) + \psi(\xi - b\sqrt{\beta} ) \right].
\end{equation}

%%%%%%%%%%%%%%%%%%%%%%%%%%%%%%%%%%%%%%%%%%%%%%%%%%%%%%%%%%%%%%%%%%%%%%%%%%%%%%
%%%%%%%%%%%%%%%%%%%%%  END MINIMAL-LENGTH SCENARIO  %%%%%%%%%%%%%%%%%%%%%%%%%%
%%%%%%%%%%%%%%%%%%%%%%%%%%%%%%%%%%%%%%%%%%%%%%%%%%%%%%%%%%%%%%%%%%%%%%%%%%%%%%

%%%%%%%%%%%%%%%%%%%%%%%%%%%%%%%%%%%%%%%%%%%%%%%%%%%%%%%%%%%%%%%%%%%%%%%%%%%%%%
%%%%%%%%%%%%%%%%%%%%%%%%%%%%  THE MODEL  %%%%%%%%%%%%%%%%%%%%%%%%%%%%%%%%%%%%%
%%%%%%%%%%%%%%%%%%%%%%%%%%%%%%%%%%%%%%%%%%%%%%%%%%%%%%%%%%%%%%%%%%%%%%%%%%%%%%
\section{The Model}
\label{Model}
%%%%%%%%%%%%%%%%%%%%%%%%%%%%%%%%%%%%%%%%%%%%%%%%%%%%%%%%%%%%%%%%%%%%%%%%%%%%%%

\h We will now consider the de Sitter Cosmology Model, which describes the phase of the Universe with an exponential expansion, during which the vacuum energy dominates the energy density and gives rise the term corresponding to the cosmological constant $\Lambda$.

%%%%%%%%%%%%%%%%%%%%%%%%%%%%%%%%%%%%%%%%%%%%%%%%%%%%%%%%%%%%%%
%%%%%%%%%%%%%%%%%  GRAVITY ACTION  %%%%%%%%%%%%%%%%%%%%%%%%%%%
%%%%%%%%%%%%%%%%%%%%%%%%%%%%%%%%%%%%%%%%%%%%%%%%%%%%%%%%%%%%%%
\subsection{Gravity Action}
\label{GravAct}
%%%%%%%%%%%%%%%%%%%%%%%%%%%%%%%%%%%%%%%%%%%%%%%%%%%%%%%%%%%%%%

The more general gravity action leading up to second order differential equations is given by,\footnote{In units so that $c =1$ and $\frac{16 \pi G}{c^{4}} = 1 $.}
\begin{equation}
	S_{g} = \int_{M} d^{4}x\sqrt{-g} (R - 2 \Lambda) + 2 \int_{\partial M} d^{3}x \sqrt{h} K,
\end{equation}
where $R$ is the Ricci curvature scalar, $K$ is the trace of the extrinsic curvature $K_{ij}$, $g$ is the determinant of the metric $g_{\mu \nu}$, $h$ is the determinant of the induced metric over the three-dimensional spatial hypersurface and $\partial M$ is the bounded of the four-dimensional manifold  $M$. The cosmological constant $\Lambda$ appears as a geometric term and its value is arbitrary. In fact, the first integral in the l.h.s. of (21) constitutes the most general  geometrical Lagrangian
in four dimensions leading to second order differential equations according to the Lovelock theorem \cite{Lovelock:1971}.

According to the Cosmological Principle, that is, for a homogeneous and isotropic universe, the metric is that of FLRW (Friedmann-Lema\^itre-Robertson-Walker),
\begin{equation}
\label{FLRW-M}
ds^{2} = -N^{2}(t)dt^{2} +
\end{equation}
$$
	 a^{2}(t) \left[ \frac{dr^{2}}{\left( 1 - kr^{2} \right)} + r^{2} d\theta^{2} +r^{2} \sin^{2}\theta d\phi^{2} \right],
$$
where $N(t)$ is the lapse function, $k = -1, 0, +1$ for hyperbolic open, flat open and close universe, respectively, and $a(t)$ is the scale factor of universe.

Using the FLRW metric, Eq.~(\ref{FLRW-M}), the gravity action becomes
\begin{equation}
	S_{g} = \int dt \left( - \frac{6 \dot{a}^{2}a}{N} + kNa - 2Na^{3}\Lambda \right).
\end{equation}
Since $S_{g} = \int dt L_{g}$ we have that the gravity Lagrangian is given by
\begin{equation}
	L_{g} =  - \frac{6 \dot{a}^{2}a}{N} + kNa - 2Na^{3}\Lambda.
\end{equation}
Hence the gravity Hamiltonian, $NH_{g} = \dot{a}p_{a} - L_{g}$, reads
\begin{equation}
	\label{hg}
	H_{g} = - \frac{p_{a}^{2}}{24a} - ka +2\Lambda a^{3},
\end{equation}
where
\begin{equation}
	p_{a} = \frac{\partial L_{g}}{\partial \dot{a}} = - \frac{12 \dot{a}a}{N}.
\end{equation}

%%%%%%%%%%%%%%%%%%%%%%%%%%%%%%%%%%%%%%%%%%%%%%%%%%%%%%%%%%%%%%
%%%%%%%%%%%%%%%%%  FLUID ACTION  %%%%%%%%%%%%%%%%%%%%%%%%%%%%%
%%%%%%%%%%%%%%%%%%%%%%%%%%%%%%%%%%%%%%%%%%%%%%%%%%%%%%%%%%%%%%
\subsection{Fluid Action}
\label{FluidAct}
%%%%%%%%%%%%%%%%%%%%%%%%%%%%%%%%%%%%%%%%%%%%%%%%%%%%%%%%%%%%%%

\h In our model the Universe is fully composed of perfect fluid. Now, we will employ the Schutz's formalism \cite{Schutz1:1970,Schutz2:1970} for the dynamics description the fluid interacting with the gravitational field \cite{Flavio:1998}. For the fluid which the state equation is,
\begin{equation}
	p = \alpha \rho,
\end{equation}
where $\rho$ is the density and $\alpha$ is the equation of state parameter depending on the type of fluid, the pressure is given by,
\begin{equation}
	\label{pressao}
	p = \alpha \left( \frac{\mu}{\alpha + 1} \right)^{1 + \frac{1}{\alpha}} e^{-\frac{S}{\alpha}},
\end{equation}
where $\mu$ is the specific enthalpy and $S$ is the entropy.

In the Schutz's formalism the four-velocity is defined by means of six potentials:
\begin{equation}
	U_{\nu} := \frac{1}{\mu} \left( \phi,_{\nu} + \epsilon \eta,_{\nu} + \theta S,_{\nu} \right).
\end{equation}
Since $\epsilon$ and $\eta$ are associated with rotation movement, they vanish. From the normalization condition,
\begin{equation}
U^{\nu}U_{\nu} = - 1,	
\end{equation}
we obtain,
\begin{equation}
	\label{entalpia}
	\mu = \frac{1}{N} \left( \dot{\phi} + \theta \dot{S} \right).
\end{equation}

The fluid action is given by
\begin{equation}
	S_{F} = \int_{M} d^{4}x \sqrt{-g}p.
\end{equation}
Using Eq.~(\ref{entalpia}) into Eq.~(\ref{pressao}) and since the spatial section is homogeneous and isotropic, we have,
\begin{equation}
	S_{F} = \int dt N a^{3} \alpha \left[ \frac{\dot{\phi} + \theta \dot{S}}{N ( \alpha + 1 )} \right]^{1 + \frac{1}{\alpha}} e^{-\frac{S}{\alpha}}.
\end{equation}
Thus, the fluid Lagrangian is,
\begin{equation}
	L_{F} = \frac{\alpha a^{3}}{N^{\frac{1}{\alpha}} (1 + \alpha)^{1 + \frac{1}{\alpha}}} \left( \dot{\phi} + \theta \dot{S} \right)^{1 + 				\frac{1}{\alpha}} e^{- \frac{S}{\alpha}}.
\end{equation}

The conjugate canonical momenta to $\phi$ and $S$ are,
\begin{equation}
	p_{\phi} = \frac{\partial L_{F}}{\partial \dot{\phi}} = \frac{ a^{3}\left( \dot{\phi} + \theta \dot{S} \right)^{\frac{1}{\alpha}} e^{- \frac{S}{\alpha}}}{N^{\frac{1}{\alpha}} (1 + \alpha)^{\frac{1}{\alpha}}}
\end{equation}
and
\begin{equation}
	\label{hf}
	p_{S} = \frac{\partial L_{F}}{\partial \dot{S}} = \theta p_{\phi}. 
\end{equation}

Therefore, the fluid Hamiltonian, $ NH_{F} = \dot{\phi} p_{\phi} + \dot{S} p_{S} - L_{F}$, is given by,
\begin{equation}
	H_{F} = \frac{p_{\phi}^{1 + \alpha} e^{S}}{a^{3 \alpha}}.
\end{equation}

From Eqs. (\ref{hf}) and (\ref{hg}), it follows that the total Hamiltonian, $H = H_{g} + H_{F}$, is
\begin{equation}
	H =  - \frac{p_{a}^{2}}{24a} - ka +2\Lambda a^{3} + \frac{p_{\phi}^{1 + \alpha} e^{S}}{a^{3 \alpha}}.
\end{equation}

Now, performing the canonical transformation (re-parametrization)
\begin{equation}
	T := - p_{S} e^{-S} p_{\phi}^{-(1 + \alpha)},
\end{equation}
and
\begin{equation}
	p_{T} := p_{\phi}^{1 + \alpha} e^{S},
\end{equation}
the total Hamiltonian becomes,
\begin{equation}
	\label{TH}
	H =  - \frac{p_{a}^{2}}{24a} - ka + 2 a^{3} \Lambda + \frac{p_{T}}{a^{3 \alpha}}.
\end{equation}
Note that $p_{T}$ is linear in Eq.~(\ref{TH}). When the quantization is implemented this fact allows us to consider $t = -T$ as playing role of time and to obtain a Schroedinger-like equation.

%%%%%%%%%%%%%%%%%%%%%%%%%%%%%%%%%%%%%%%%%%%%%%%%%%%%%%%%%%%%%%%%%%%%%%%%%%%%%%
%%%%%%%%%%%%%%%%%%%%%%%%  END THE MODEL  %%%%%%%%%%%%%%%%%%%%%%%%%%%%%%%%%%%%%
%%%%%%%%%%%%%%%%%%%%%%%%%%%%%%%%%%%%%%%%%%%%%%%%%%%%%%%%%%%%%%%%%%%%%%%%%%%%%%

%%%%%%%%%%%%%%%%%%%%%%%%%%%%%%%%%%%%%%%%%%%%%%%%%%%%%%%%%%%%%%%%%%%%%%%%%%%%%%
%%%%%%%%%  THE QUANTUM COSMOLOGY MODEL IN A MINIMAL-LENGTH SCENARIO  %%%%%%%%%
%%%%%%%%%%%%%%%%%%%%%%%%%%%%%%%%%%%%%%%%%%%%%%%%%%%%%%%%%%%%%%%%%%%%%%%%%%%%%%
\section{The Quantum Cosmology Model in a Minimal-Length Scenario}
\label{QCML}
%%%%%%%%%%%%%%%%%%%%%%%%%%%%%%%%%%%%%%%%%%%%%%%%%%%%%%%%%%%%%%%%%%%%%%%%%%%%%%

%%%%%%%%%%%%%%%%%%%%%%%%%%%%%%%%%%%%%%%%%%%%%%%%%%%%%%%%%%%%%%
%%%%%%%%%%%%%%%%% THE MODIFIED WDW EQ  %%%%%%%%%%%%%%%%%%%%%%%
%%%%%%%%%%%%%%%%%%%%%%%%%%%%%%%%%%%%%%%%%%%%%%%%%%%%%%%%%%%%%%
\subsection{The modified Wheeler-DeWitt equation}
\label{mWDWe}
%%%%%%%%%%%%%%%%%%%%%%%%%%%%%%%%%%%%%%%%%%%%%%%%%%%%%%%%%%%%%%

\h The quantization process takes place in the framework of mini-superspace and according with the Wheeler-DeWitt quantization scheme,
\begin{equation}
	p_{a} \longrightarrow \hat{p}_{a}
\end{equation}
and
\begin{equation}
	p_{T} \longrightarrow \hat{p}_{T} \equiv -i\frac{\partial}{\partial T},
\end{equation}
such that $H \longrightarrow \hat{H}$, implying the WDW equation,
\begin{equation}
	\hat{H} \ket{\psi} = 0.
\end{equation}

In order to obtain a minimal-length scenario we demand that
\begin{equation}
\label{drca}
	[\hat{a},\hat{p}_{a}] := \frac{i\hbar}{1 - \beta\hat{p}_{a}^2}.
\end{equation}
with the representation of the operators as follow:
\begin{equation}
	\label{ar}
	\hat{a} = \hat{x}
\end{equation}
and
\begin{equation}
	\label{pra}
	\hat{p}_{a} \equiv \hat{p} + \frac{\beta}{3} \hat{p}^{3} +\frac{\beta^{2}}{3} \hat{p}^{5} + {\cal O} \left( \beta^{3} \right),
\end{equation}
where the $\hat{x}$ and $\hat{p}$ operators satisfy the canonical commutation relation,
\begin{equation}
\label{drcax}
	[\hat{x},\hat{p}] := i\hbar. 
\end{equation} 

Using the representation (\ref{ar}) and (\ref{pra}), the modified WDW equation turns out
$$
	- \frac{\hat{x}^{(3 \alpha -1)}}{24} \left[ \hat{p}^{2} + \frac{2}{3} \beta \hat{p}^{4} +  \frac{7}{9} \beta^{2} \hat{p}^{6} + {\cal O} \left( \beta^{3} \right) \right] \ket{\psi(t)}
$$
\begin{equation}
\label{mwdwe}
+  \left[ - k \hat{x}^{(1 + 3 \alpha)} + 2 \Lambda \hat{x}^{3(1 + \alpha)} \right]\ket{\psi(t)}  = -i\frac{\partial}{\partial t} \ket{\psi(t)},
\end{equation}
where we have made $T = -t$.

Now, projecting Eq.~(\ref{mwdwe}) onto the formal representation of the ``position'' space we have,
\begin{equation}
	\label{mwdwe}
	\langle x | \hat{H} | \psi(t) \rangle = 0,
\end{equation}
$$
	- \frac{x^{(3 \alpha -1)}}{24} \left[  - \frac{\partial^{2}}{\partial x^{2}} + \frac{2}{3} \beta \frac{\partial^{4}}{\partial x^{4}} - \frac{7}{9} \beta^{2} \frac{\partial^{6}}{\partial x^{6}} + {\cal O} \left( \beta^{3} \right) \right]\psi(x,t)
$$
\begin{equation}
	\label{mwdwex}
	+  \left[ -k x^{(1 + 3 \alpha)} + 2 \Lambda x^{3(1 + \alpha)} \right] \psi(x,t) = -i\frac{\partial}{\partial t} \psi(x,t),
\end{equation}
where we have used Eqs. (\ref{repx}) and (\ref{repp}).

Assuming that the solution of the above equation can been written as $\psi(x,t) = \varphi(x) \tau(t)$, we find the stationary solutions,
\begin{equation}
	\psi_{\omega}(x,t) = e^{-i \omega t} \varphi_{\omega}(x),
\end{equation}
and the time independent equation,
$$
	\left[ \frac{d^{2}}{dx^{2}} - \frac{2}{3} \beta \frac{d^{4}}{dx^{4}} + \frac{7}{9} \beta^{2} \frac{d^{6}}{dx^{6}} + {\cal O} \left( \beta^{3} \right) \right] \varphi_{\omega}(x)
$$
\begin{equation}
	\label{wdweix}
	 + \biggl[ 24 x^{(1 - 3 \alpha)} \omega - 24kx^{2} + 48 \Lambda x^{4} \biggr] \varphi_{\omega}(x) = 0,
\end{equation}
where $\omega$ is a constant.

The equation of state parameter $\alpha$ defines the matter content in a single fluid model.
The case $\alpha = - 1$ corresponds to the vacuum energy equation of state. It comes from the quantum field contributions for the energy associated with the vacuum state. Its precise value is the object of discussions today, and it can be related to the dark energy component in the Universe. Here, we consider it as a dynamic fluid component with internal degrees of freedom. Remark that the contribution of this vacuum component for the matter content, given by $\alpha = - 1$, is similar to the geometric cosmological constant $\Lambda$, but the physical meaning and origin are very different.

Lastly, considering a flat universe\footnote{We choose a flat universe because it is in agreement with current cosmological observations \cite{Bianchini:2020,Efstathiou:2020}.} ($k = 0$), without any other matter content, fully completed by a dynamical vacuum with an equation of state,
\begin{equation}
	P = - \rho,
\end{equation}
and $\Lambda = 0$,\footnote{In the same way as in \cite{Flavio:1998} we hope that the cosmological constant arises from the degrees of freedom of the vacuum and not being postulated from the start as a geometric term. Note also that if from the start $\Lambda > 0$ we can always redefine $\omega$ as $\omega\prime := \omega + 2\Lambda$.} Eq.~(\ref{wdweix}) up to ${\cal O} \left( \beta^{2} \right)$ becomes,\footnote{From now on, for the sake of simplicity we are going to omit the subscript $\omega$.}
\begin{equation}
	\label{wdwe}
	\frac{d^{2}\varphi}{dx^{2}} - \frac{2}{3} \beta \frac{d^{4}\varphi}{dx^{4}} + \frac{7}{9} \beta^{2} \frac{d^{6}\varphi}{dx^{6}} +24\omega x^{4} \varphi = 0.
\end{equation}

%%%%%%%%%%%%%%%%%%%%%%%%%%%%%%%%%%%%%%%%%%%%%%%%%%%%%%%%%%%%%%
%%%%%%%%%%%%%%% SOLUTION MODIFIED WDW EQ  %%%%%%%%%%%%%%%%%%%%
%%%%%%%%%%%%%%%%%%%%%%%%%%%%%%%%%%%%%%%%%%%%%%%%%%%%%%%%%%%%%%
\subsection{Solution of the modified Wheeler-DeWitt equation}
\label{smWDWe}
%%%%%%%%%%%%%%%%%%%%%%%%%%%%%%%%%%%%%%%%%%%%%%%%%%%%%%%%%%%%%%

\h It is not possible to solve  Eq.~(\ref{wdwe}) analytically since we do not know all initial or boundary conditions\footnote{In Ref. \cite{Kober1:2012} the author has solved a modified WDW equation of forth order $\left( \frac{d^{4}}{dx^{4}} \right)$ using the Sommerfeld polynomial method. However, the solution holds dependent on four parameters that must be determined by the initial conditions, which are not known.}, that is, $\varphi^{(N)}(x_{0})$, for $ N = 0, 1, ..., 5$. A way to get around this problem is considering $x$ small. This is well reasonable since we are interested in solutions describing the Universe in its quantum regime, that is, in the initial phases of the Universe, when the scale factor was small\footnote{Later on, we will discuss more carefully the issue if $x$ can rigorously describe the scale factor.} and as it is known quantum effects are significant only for small values of the scale factor.

Initially, we take into account Eq.~(\ref{wdwe}) in the limit $\beta = 0$, which we call ordinary WDW equation,
\begin{equation}
	\label{owdwe}
	\frac{d^{2}\varphi_{0}}{dx^{2}} +24\omega x^{4} \varphi_{0} = 0.
\end{equation}
Making $\varphi_{0} = \sqrt{x}y(x)$ and the change of variable $\rho := \sqrt{\frac{8 \omega}{3}} x^{3}$, Eq.~(\ref{wdwe}) turns the Bessel equation,
\begin{equation}
	\label{be}
	\frac{d^{2}y}{d \rho^{2}} + \frac{1}{\rho}\frac{dy}{d \rho} + \left[ 1 - \frac{\left( \frac{1}{6} \right)^{2}}{\rho^{2}} \right] y = 0.
\end{equation}
Then, the solution of Eq.~(\ref{be}) is
\begin{equation}
	\varphi_{0}(x) = A \sqrt{x} J_{1/6} \left( K_{0} x^{3} \right) + B \sqrt{x} N_{1/6} \left( K_{0}x^{3} \right),
\end{equation}
where $K_{0} := \sqrt{\frac{8 \omega}{3}}$, $J_{1/6}(X)$ and $N_{1/6}(X)$ are the Bessel and the Neumann functions of order $1/6$, respectively, and $A$ and $B$ are constants.

It is easy to see that in the limit for small $x$,
\begin{equation}
	J_{1/6}(X) \approx \sqrt{x}
\end{equation}
and
\begin{equation}
	N_{1/6}(X) \approx \frac{1}{\sqrt{x}}.
\end{equation}

Now, imposing the condition\footnote{That condition is obtained by demanding that the Hamiltonian $H$ must be self-adjoint \cite{Flavio:1998}. That condition should not change, even though it is probably that boundary conditions satisfied by derivatives change. Note that condition does not avoid the initial singularity \cite{Battisti1:2008,Battisti2:2008}.} $\varphi_{0}(0) = 0$, we get $B = 0$, and the solution is given by,
\begin{equation}
	\varphi_{0}(x) = A \sqrt{x} J_{1/6} \left( K_{0} x^{3} \right).
\end{equation}

Remembering that
\begin{equation}
	J_{\nu}(X) = \sum_{N=0}^{\infty} \frac{(-1)^{N}}{N! \Gamma(N + \nu + 1)} \left( \frac{X}{2} \right)^{2N + \nu},
\end{equation}
and retaining the first three significant terms of $\varphi_{0}(x)$ we get,
\begin{equation}
	\label{as1}
	\varphi_{0}(x) = C_{1}x + C_{2}x^{7} + C_{3}x^{13}.
\end{equation}
It is appropriate to rewrite the solution (\ref{as1}) by absorbing $C_{1}$ in a normalization constant, which we are going to omit without affecting the results:
\begin{equation}
	\label{as2}
	\varphi_{0}(x) = x - \frac{4}{7} \omega x^{7} + \frac{8}{91} \omega^{2} x^{13}.
\end{equation}

We can obtain an approximate solution for Eq.~(\ref{wdwe}), in the range of $x$ small, using Eq.~(\ref{as2}) into  ${\cal O}\left( \beta \right)$ and ${\cal O}\left( \beta^{2} \right)$ terms. Thus,
\begin{equation}
	\label{d4}
	\frac{d^{4} \varphi_{0}}{dx^{4}} = -480 \omega x^{3} + \frac{10560}{7} \omega^{2} x^{9}
\end{equation}
and
\begin{equation}
	\label{d6}
	\frac{d^{6} \varphi_{0}}{dx^{6}} = -2880 \omega x + \frac{760320}{7} \omega^{2} x^{7}.
\end{equation}
With the approximation $\varphi \approx x$ into Eqs.~(\ref{d4}) and (\ref{d6}), Eq.~(\ref{wdwe}) becomes,
$$
	\frac{d^{2} \varphi}{dx^{2}} +24 \omega x^{4} \varphi + \beta \left( 320 \omega x^{2} \varphi - \frac{7040}{7} \omega^{2} x^{8} \varphi \right) +
$$
\begin{equation}
	\label{aeq1}
 \beta^{2} \left( -2240 \omega \varphi + 591360 \omega^{2} x^{6} \varphi \right) = 0.
\end{equation}

Subsequently, we are going to find the approximate solution of the above equation in two different ways.

%%%%%%%%%%%%%%%%%%%%%%%%%%%%%%%%%%%%%%%%%%%%%%%%%%%%%%%%%%%%%%
%%%%%%%%%%%%%%% SOLUTION MODIFIED WDW EQ  %%%%%%%%%%%%%%%%%%%%
%%%%%%%%%%%%%%%%%%%%%%%%%%%%%%%%%%%%%%%%%%%%%%%%%%%%%%%%%%%%%%
\subsubsection{Solution: first method}
\label{s1m}
%%%%%%%%%%%%%%%%%%%%%%%%%%%%%%%%%%%%%%%%%%%%%%%%%%%%%%%%%%%%%%

\h Since $x = 0$ is an ordinary point of Eq.~(\ref{aeq1}) we can find a solution in power series of $x$. Then, replacing $ \varphi(x) = \sum_{n=0}^{\infty} a_{n}x^{n}$ into Eq.~(\ref{aeq1}) and retaining only significant terms, we have,
$$
		\varphi(x) = x - \frac{4}{7} \omega x^{7} + \frac{8}{91} \omega^{2} x^{13} + 
$$
\begin{equation}
	\label{sol1}
\beta \left( - 16 A \omega x^{5} + \frac{5504}{385} B \omega^{2} x^{11} \right) + \beta^{2} \frac{1120}{3} C \omega x^{3}.
\end{equation}
The parameters $A$, $B$ and $C$ are determined by requiring that~(\ref{sol1}) satisfies Eq.~(\ref{aeq1}) up to order chosen in $x$. Then, substituting (\ref{sol1})~into~(\ref{aeq1}) we have $A = 1$, $B = \frac{38}{43}$  and $C = \frac{3}{7}$. Therefore, the solution reads as,
$$
		\varphi(x) = x - \frac{4}{7} \omega x^{7} + \frac{8}{91} \omega^{2} x^{13} + 
$$
\begin{equation}
	\label{sol2}		
		\beta \left( - 16 \omega x^{5} + \frac{4864}{385} \omega^{2} x^{11} \right) + \beta^{2} 160 \omega x^{3}.
\end{equation}

%%%%%%%%%%%%%%%%%%%%%%%%%%%%%%%%%%%%%%%%%%%%%%%%%%%%%%%%%%%%%%
%%%%%%%%%%%%%%% SOLUTION MODIFIED WDW EQ  %%%%%%%%%%%%%%%%%%%%
%%%%%%%%%%%%%%%%%%%%%%%%%%%%%%%%%%%%%%%%%%%%%%%%%%%%%%%%%%%%%%
\subsubsection{Solution: second method}
\label{s2m}
%%%%%%%%%%%%%%%%%%%%%%%%%%%%%%%%%%%%%%%%%%%%%%%%%%%%%%%%%%%%%%

\h The same previous solution can be obtained by taking the approximation following,
\begin{equation}
	\label{aeq2}
	\frac{d^{2}\varphi}{dx^{2}} = - 24\omega x^{4} \varphi_{0} + \frac{2}{3} \beta \frac{d^{4}\varphi_{0}}{dx^{4}} - \frac{7}{9} \beta^{2} \frac{d^{6}\varphi_{0}}{dx^{6}}.
\end{equation}

Integrating twice the above equation we obtain,
$$
	\frac{d^{2}\varphi}{dx^{2}} = - 24\omega \int dx \int dx x^{4} \varphi_{0} +
$$
\begin{equation}
	\label{int2}
 \frac{2}{3} \beta \frac{d^{2}\varphi_{0}}{dx^{2}} - \frac{7}{9} \beta^{2} \frac{d^{4}\varphi_{0}}{dx^{4}} + \alpha_{1} x + \alpha_{2},
\end{equation}
where $\alpha_{1}$ and $\alpha_{2}$ are integration constants.
After that, using~(\ref{as2}) into~(\ref{int2}) we obtain,
$$
		\varphi(x) = \alpha_{2} + \alpha_{1} x  - \frac{4}{7} \omega x^{7} + \frac{8}{91} \omega^{2} x^{13} + 
$$
\begin{equation}
	\label{sol3}
\beta \left( - 16 A' \omega x^{5} + \frac{64}{7} B' \omega^{2} x^{11} \right) + \beta^{2} \frac{120}{3} C' \omega x^{3}.
\end{equation}
The constants $\alpha_{1}$ and $\alpha_{2}$ can be determined by demanding that $\varphi(x) \longrightarrow \varphi_{0}(x)$ when $\beta \longrightarrow 0$. In a similar way, the parameters $A'$, $B'$ and $C'$ are found requiring that the solution~(\ref{sol3}) satisfies Eq.~(\ref{aeq1}). Consequently, we again have the solution (\ref{sol2}), as we expected.

%%%%%%%%%%%%%%%%%%%%%%%%%%%%%%%%%%%%%%%%%%%%%%%%%%%%%%%%%%%%%%
%%%%%%%%%%%%%%% VALIDITY RANGE  %%%%%%%%%%%%%%%%%%%%
%%%%%%%%%%%%%%%%%%%%%%%%%%%%%%%%%%%%%%%%%%%%%%%%%%%%%%%%%%%%%%
\subsection{Validity range}
\label{vr}
%%%%%%%%%%%%%%%%%%%%%%%%%%%%%%%%%%%%%%%%%%%%%%%%%%%%%%%%%%%%%%

\h In order that the solution (\ref{sol2}) works consistently the first disregarded term in $\varphi_{0}(x)$ must be much smaller than the last kept terms in ${\cal O}(\beta)$ and ${\cal O} \left( \beta^{2} \right)$, that is, $ x \ll \left( \frac{\beta}{\omega} \right)^{\frac{1}{8}}$. Since $\beta$ is much small, this drastically reduces the validity range for our solution. We can improve this if terms of higher powers in $x$ are taken account in order ${\cal O}\left( \beta^{0} \right)$ part of the solution. Therefore we can consider $\sqrt{x}J_{1/6} \left( K_{0} x^{3} \right)$ as the part in ${\cal O}\left( \beta^{0} \right)$ of $\varphi(x)$, that is, $\varphi_{0}(x)$. It is clear that terms of order such that $x^{6n - 2} < \beta^{2} \omega^{1 - n}$ lie far outside the validity range and they must throw away.

The same reasoning applies in the case of terms in ${\cal O}\left( \beta^{2} \right)$ when compared to terms in ${\cal O}(\beta)$. This way, with the purpose of increasing the validity range of our solution we take account the two first significant terms in ${\cal O}(\beta)$. In conclusion, in light of the forgoing the solution turns out to be,\footnote{Note that solution is non-normalized.}
$$
		\varphi(x) = \sqrt{x}J_{1/6} \left( K_{0} x^{3} \right) + 
$$
\begin{equation}
	\label{sol4}
\beta \left( - 16 \omega x^{5} + \frac{4864}{385} \omega^{2} x^{11} \right) + \beta^{2} 160 \omega x^{3}.
\end{equation}

In dealing with expansions of small parameters (in this case $\beta$) it is necessary to take care to consistently work when $x$ goes to an extremely small value. It is easy to see that terms in $\beta^{0}$ are like $\omega^{N} x^{1 + 6N}$, terms in $\beta$ are like $ \beta \omega^{N} x^{5 + 6N}$ and terms in $\beta^{2}$ are like $ \beta^{2} \omega^{N} x^{3 + 6N}$. Therefore, the terms in $\beta \omega x^{5}$ and $\beta \omega^{2} x^{11}$ are negligible when $x < \sqrt{\beta}$ and the term $ \beta \omega^{2} x^{11}$ is negligible when $\sqrt{\beta} < x < \beta^{1/8}$. Note that the term in $\beta^{0} x$ always will be greater than the term in $\beta^{2} \omega x^{3}$ for $x$ small.

Keep in mind that in fact solution (\ref{sol4}) is $\varphi_{\omega}(x)$, that is, eigenfunctions whose associated eigenvalues are values of the cosmology constant.

%%%%%%%%%%%%%%%%%%%%%%%%%%%%%%%%%%%%%%%%%%%%%%%%%%%%%%%%%%%%%%
%%%%%%%%%%%%% PHYSICALLY ACCEPTABLE SOLUTIONS  %%%%%%%%%%%%%%%
%%%%%%%%%%%%%%%%%%%%%%%%%%%%%%%%%%%%%%%%%%%%%%%%%%%%%%%%%%%%%%
\subsection{Physically acceptable solutions}
\label{pas}
%%%%%%%%%%%%%%%%%%%%%%%%%%%%%%%%%%%%%%%%%%%%%%%%%%%%%%%%%%%%%%

\h As we have already said, we can not obtain directly from $\varphi(x)$ physical results because the $\hat{x}$ operator eigenstates do not belong to the Hibert space. Nevertheless, the projections of the state vectors $\ket{\varphi}$ on the maximal localization states, that is, the wave functions in the representation of the quasi-position space do. The wave functions in the quasi-position space can easily be obtained from (\ref{sol4}) by using Eq.~(\ref{wfqp}). Therefore, replacing (\ref{sol4}) into (\ref{wfqp}) we have\footnote{Remember that $b := \frac{3 \pi}{4}$, Eq.~(\ref{b}).}
$$
	\varphi_{qp}(\xi) = \sqrt{\xi}J_{1/6} \left( K_{0} \xi^{3} \right) +
$$
$$
\beta \left[ - \left( 16 + 12 b^{2} \right) \omega \xi^{5} + \left( \frac{4864}{385} + \frac{48}{7} b^{2} \right) \omega^{2} \xi^{11} \right] +
$$
\begin{equation}
	\label{sol5}
 \beta^{2} \left( 160 - 160 b^{2} - 20 b^{4} \right) \omega \xi^{3}.
\end{equation}

Hence, above quasi-position wave function represents the probability amplitude for the Universe being maximally localized around the position $\xi$.

The reader may be questioning if the correct procedure would be to determine the solution of the modified WDW equation in the representation of the quasi-position space,
\begin{equation}
	\label{qprepeq}
	\langle \psi_{\xi}^{ml} | \left( - \hat{p}_{a}^{2} + 24 \omega \hat{a}^{4} \right) | \psi(t) \rangle = 0.
\end{equation}

Using Eqs.~(\ref{qprepX}) and (\ref{qprepP}) we obtain
$$
	\frac{d^{2} \varphi_{qp}}{d \xi^{2}} +24 \omega \xi^{4} \varphi_{qp} +
$$
$$
	\beta b^{2} \left[ - \frac{2}{3 b^{2}} \frac{d^{4} \varphi_{qp}}{d \xi^{4}} + 144 \omega \xi^{2} \varphi_{qp} + 96 \omega \xi^{3} \frac{d \varphi_{qp}}{d \xi} \right] +
$$
$$
	\beta^{2} b^{4} \left[ - \frac{7}{9 b^{4}} \frac{d^{6} \varphi_{qp}}{d \xi^{6}} + 24 \omega \varphi_{qp} + 96 \omega \xi \frac{d \varphi_{qp}}{d \xi} \right] +
$$
\begin{equation}
	\label{qprepeq1}
\beta^{2} b^{4} \left[ - 24 \omega \xi^{2}  \frac{d^{2} \varphi_{qp}}{d \xi^{2}} - 40 \omega \xi^{3}  \frac{d^{3} \varphi_{qp}}{d \xi^{3}} \right] = 0.
\end{equation}
As it can easily be checked (after some algebra) the solution~(\ref{sol5}) satisfies Eq.~(\ref{qprepeq1}) up to the considered validity order.

%%%%%%%%%%%%%%%%%%%%%%%%%%%%%%%%%%%%%%%%%%%%%%%%%%%%%%%%%%%%%%
%%%%%%%%%%%%% NORM OF THE WAVE FUNCTION  %%%%%%%%%%%%%%%%%%%%%
%%%%%%%%%%%%%%%%%%%%%%%%%%%%%%%%%%%%%%%%%%%%%%%%%%%%%%%%%%%%%%
\subsection{Norm of the Wave Function}
\label{nwf}
%%%%%%%%%%%%%%%%%%%%%%%%%%%%%%%%%%%%%%%%%%%%%%%%%%%%%%%%%%%%%%

\h In the ordinary case the norm of the wave function solution of the WDW equation is infinite \cite{Flavio:1998}. This means that stationary states of the Universe with well-defined values of the cosmological constant are not realizable and it is necessary the construction of wave packets. 

Now, we are going to calculate the norm of the wave function (\ref{sol5}). Since the determination of the norm of the wave function in the quasi-position space can be laborious, it is convenient to perform first the calculation formally in the ``position'' space and after to obtain the normalized wave function (if it is normalized) in the quasi-position space using Eq.~(\ref{wfqp}).

In the ``position'' space the inner product does not change, so
\begin{equation}
	\label{ipps}
	\bkt{\psi_{1}}{\psi_{2}} = \int_{0}^{\infty} \psi_{2}^{*}(x) \psi_{1}(x) dx.
\end{equation}
On the basis of solution~(\ref{sol2}) we assume that $\psi_{1}^{(n)}(0) = 0$ for $n = 0, 2, 4, \dots$. Thus, it is easy to show that the operator,
\begin{equation}
	\label{Dop}
	\hat{D} := \frac{d^{2}}{dx^{2}} - \frac{2}{3} \beta \frac{d^{4}}{dx^{4}} + \frac{7}{9} \beta^{2} \frac{d^{6}}{dx^{6}}
\end{equation}
is self-adjoint\footnote{Note that $\hat{D}$ is not defined on the whole Hilbert space ${\cal H}$ because $\hat{D}$ is an unbounded linear operator. Thus, the domain of $\hat{D}$ is a dense subspace into ${\cal H}$.}.

However, the operator $\hat{H} := \hat{D} +24 \omega x^{4}$ is not self-adjoint in the inner product~(\ref{ipps}), but rather in the inner product
\begin{equation}
	\label{ipps2}
	\bkt{\psi_{1}}{\psi_{2}} = \int_{0}^{\infty} x^{4} \psi_{2}^{*}(x) \psi_{1}(x) dx,
\end{equation}
where $x^{4} = w(x)$ is a weight function.

Some caution is necessary with the calculation of the norm of the solution~(\ref{sol4}) because the solution~(\ref{sol4}) is only valid for small values of $x$. Therefore it can not be used throughout the whole positive $x$-axis. Thus, we consider that the effects of the existence of a minimal length are negligible for values of $x$ greater than those of the validity range of the solution~(\ref{sol4}), say $x_{M}$. Consequently, the solution turns to the ordinary $\varphi_{0}(x)$ for $x > x_{M}$. In this way,
\begin{equation}
	\label{ipps3}
	\bkt{\varphi}{\varphi} = \int_{0}^{x_{M}} x^{4} \mid \varphi(x) \mid^{2} dx +  \int_{x_{M}}^{\infty} x^{4} \mid \varphi_{0}(x) \mid^{2} dx.
\end{equation}
Note that if $x_{M} \approx \left( \frac{\beta}{\omega} \right)^{\frac{1}{8}}$ then $\varphi(x)$ of the first integral in the l.h.s. of (\ref{ipps3}) is given by Eq.~(\ref{sol2}), according to subsection \ref{vr}. Because of the second integral in the l.h.s. of (\ref{ipps3}) the norm is infinite. This result was already expected since the divergence takes place for $x \rightarrow \infty$ where the effects of a minimal length are negligible. In the ordinary case, the problem with the infinite norm is solved by constructing a wave packet. The construction of a wave packet when the minimal length is presented is much less simple due to the modifications given by the existence of a minimal length near the origin.

Although the wave function (\ref{sol5}) can not be normalized, it is interesting to note that the leading correction term in ${\cal O}(\beta)$ of $| \varphi_{qp}(\xi) |^{2}$ is
$$
-2\beta \left( 16 + 12b^{2} \right)\omega\xi^{6}.$$
Consequently,  $| \varphi_{qp} |^{2}$ grows more slowly than  $| \varphi_{0} |^{2}$. In \cite{Vakili:2009} the author\footnote{The author has considered that the variable $u := a^{\frac{2}{3}}$ and its conjugate momentum $p_{u} = \frac{2}{3} a^{-\frac{1}{2}} p_{a}$ satisfy the commutation relation proposed by Kempf, Mangano and Mann \cite{Kempf1:1995} when turned operators.} has found that the leading correction term is positive implying a more rapid growth of $| \varphi |^{2}$.

%%%%%%%%%%%%%%%%%%%%%%%%%%%%%%%%%%%%%%%%%%%%%%%%%%%%%%%%%%%%%%%%%%%%%%%%%%%%%%
%%%%%%%%% END NORM OF THE WAVE FUNCTION  %%%%%%%%%%%%%%%%%%%%%%%%%%%%%%%%%%%%%
%%%%%%%%%%%%%%%%%%%%%%%%%%%%%%%%%%%%%%%%%%%%%%%%%%%%%%%%%%%%%%%%%%%%%%%%%%%%%%

%%%%%%%%%%%%%%%%%%%%%%%%%%%%%%%%%%%%%%%%%%%%%%%%%%%%%%%%%%%%%%%%%%%%%%%%%%%%%%
%%%%%%%%% END THE QUANTUM COSMOLOGY MODEL IN A MINIMAL-LENGTH SCENARIO  %%%%%%
%%%%%%%%%%%%%%%%%%%%%%%%%%%%%%%%%%%%%%%%%%%%%%%%%%%%%%%%%%%%%%%%%%%%%%%%%%%%%%

%%%%%%%%%%%%%%%%%%%%%%%%%%%%%%%%%%%%%%%%%%%%%%%%%%%%%%%%%%%%%%%%%%%%%%%%%%%%%%%%%%%%%%%%%%%%%%%%%%%
%%%%%%%%%%%%%%%%%%%%%%%%%%%%%% CONCLUSION %%%%%%%%%%%%%%%%%%%%%%%%%%%%%%%%%%%%%%%%%%%%%%%%%%%%%%%%%
%%%%%%%%%%%%%%%%%%%%%%%%%%%%%%%%%%%%%%%%%%%%%%%%%%%%%%%%%%%%%%%%%%%%%%%%%%%%%%%%%%%%%%%%%%%%%%%%%%%
\section{Conclusion}
\label{Concl}

\h In this work, we performed the study of the primordial Universe by implementing a quantum approach in the gravitation effects. It follows the need to implement a minimal-length scenario, which is carried out by using a GUP. Since we chose the GUP (\ref{gup}) we can expand the representation of the momentum until ${\cal O}(\beta^{2})$ and thus to obtain a modified WDW equation up to ${\cal O}(\beta^{2})$, too.

We found the modified WDW equation in the formal representation of ``position'' space, Eq.~(\ref{wdwe}), and its solution $\varphi(x)$, Eq.~(\ref{sol4}), because it is simpler than in the representation of quasi-position space. However, we can not obtain directly from $\varphi(x)$ physical results. Consequently we obtained the wave function of the Universe in the representation of quasi-position space, $\varphi_{qp}(\xi)$, as a superposition of two wave functions of the Universe in the formal representation of ``position'' space, $\varphi_{qp}(\xi) = \frac{1}{\sqrt{2}}\left[\varphi(\xi + x_{min}) + \varphi(\xi - x_{min}) \right]$. With the aim of insuring our result, we found the modified WDW equation in the representation of quasi-position space, Eq.~(\ref{qprepeq1}), and we checked that $\varphi_{qp}(\xi)$ is actually its solution.

The ignorance of the initial or boundary conditions, that is, of the derivatives of the wave function does not allow us to find an exact solution for modified WDW equation, what forced us to seek a solution for small values of the scale factor of the Universe.

Since in the ordinary case the divergence in the norm of the wave function takes place for $x \rightarrow \infty$, the introduction of a minimal length in the theory can not change this, because its effects are only significant for small $x$. Therefore realizable states of the Universe are wave packets constructed by superposition of the stationary states of different values of $\omega$, that is, of the cosmological constant.

We should note that different changes in the WDW equation due to different GUP's have been found in the literature. For example, a linear GUP \cite{Ali:2009} has led to the emergence of third and fourth order derivatives in the modified WDW equation \cite{Kober1:2012,Faizal:2014,Faizal:2016,Majumder1:2011}, whereas the KMM GUP \cite{Kempf1:1995} has only induced to fourth order derivatives \cite{Vakili:2007,Faizal:2015,Vakili:2009,Zeynali:2012}. Those results are in agreement with our modified WDW equation which also displays a term of fourth order derivative in ${\cal O}(\beta)$.

Fluid radiation is another model of interest to study since the radiation-dominated era has been presented in the early Universe when the quantum effects were relevant. For a radiation fluid $\alpha = \frac{1}{3}$ and the modified WDW equation is given by
\begin{equation}
	\label{wdwerad}
	\frac{d^{2}\varphi}{dx^{2}} - \frac{2}{3} \beta \frac{d^{4}\varphi}{dx^{4}} + \frac{7}{9} \beta^{2} \frac{d^{6}\varphi}{dx^{6}} + 24\omega\varphi = 0.
\end{equation} 
At first sight we could easily obtain an analytical solution of the above equation. For this we suppose $\varphi(x) = e^{k x}$ in order to get an algebraic equation of 6th order which can be transformed in one of 3rd order. However, we do not know the initial or boundary conditions in order to determine the 6 constants of the general solution. Again, we need to appeal to the approximation for small $x$.

In next works, we are going to study the effects on the scale factor evolution, applications of the obtained results and comparisons with others models and GUP's.

%%%%%%%%%%%%%%%%%%%%%%%%%%%%%%%%%%%%%%%%%%%%%%%%%%%%%%%%%%%%%%%%%%%%%%%%%%%%%%%%%%%%%%%%%%%%%%%%%%%
%%%%%%%%%%%%%%%%%%%%%%%%%%%%%% END Clonclusion %%%%%%%%%%%%%%%%%%%%%%%%%%%%%%%%%%%%%%%%%%%%%%%%%%%%
%%%%%%%%%%%%%%%%%%%%%%%%%%%%%%%%%%%%%%%%%%%%%%%%%%%%%%%%%%%%%%%%%%%%%%%%%%%%%%%%%%%%%%%%%%%%%%%%%%%

%%%%%%%%%%%%%%%%%%%%%%%%%%%%%%%%%%%%%%%%%%%%%%%%%%%%%%%%%%%%%%%%%%%%%%%%%%%%%%%%%%%%%%%%%%%%%%%%%%%
%%%%%%%%%%%%%%%%%%%%%%%%%%%%%%%%%%%%%%%%%%%%%%%%%%%%%%%%%%%%%%%%%%%%%%%%%%%%%%%%%%%%%%%%%%%%%%%%%%%
\section*{Acknowledgements}

\h We would like to thank FAPES, CAPES and CNPq (Brazil) for financial support.
\\

%%%%%%%%%%%%%%%%%%%%%%%%%%%%%%%%%%%%%%%%%%%%%%%%%%%%%%%%%%%%%%%%%%%%%%%%%%%%%%%%%%%%%%%%%%%%%%%%%%%%%%%%%%%%%%%%%%%%%%%%%%%%%%%%%%%%%%%%%%%%%%%%%%%%%%%%%%%%%%%%%%%%%%%%%%%%%%%%%%%%%%%%%%%%%%%%%%%%%%

%\newpage

%%%%%%%%%%%%%%%%%%%%%%%%%%%%%%%%%%%%%%%%%%%%%%%%%%%%%%%%%%%%%%%%%%%%%%%%%%%%%%%%%%%%%%%%%%%%%%%%%%%%%
%%%%%%%%%%%%%%%%%%%%%%%%%%%%%% REFERENCES %%%%%%%%%%%%%%%%%%%%%%%%%%%%%%%%%%%%%%%%%%%%%%%%%%%%%%%%%
%%%%%%%%%%%%%%%%%%%%%%%%%%%%%%%%%%%%%%%%%%%%%%%%%%%%%%%%%%%%%%%%%%%%%%%%%%%%%%%%%%%%%%%%%%%%%%%%%%%
%\section*{References}

%\bibliography{mybibfile}
%\begin{thebibliography}{000} %for 3 digits

\end{document}